\documentstyle[12pt]{article}

\newcommand{\be}{\begin{equation}}
\newcommand{\ee}{\end{equation}}
\begin{document}


\begin{center}
Journal of Physics: Conference Series 7 (2005) 17-33 
\end{center}

\begin{center}

{\Large \bf Fractional Liouville and BBGKI Equations}

\vskip 7mm
{\large \bf Vasily E. Tarasov }

{\it Skobeltsyn Institute of Nuclear Physics, \\
Moscow State University, Moscow 119992, Russia}

E-mail: tarasov@theory.sinp.msu.ru 
\end{center}

\begin{abstract}
We consider the fractional generalizations of Liouville equation. 
The normalization condition, phase volume, and average values are 
generalized for fractional case.
The interpretation of fractional analog of phase space as 
a space with fractal dimension and as a space with fractional measure 
are discussed. The fractional analogs of the Hamiltonian systems 
are considered as a special class of non-Hamiltonian systems.
The fractional generalization of the 
reduced distribution functions are suggested.
The fractional analogs of the BBGKI 
equations are derived from the fractional Liouville equation.
\end{abstract}

\section{Introduction}

Fractional integrals and derivatives have many
applications in statistical mechanics and kinetics \cite{Zas2}. 
G.M. Zaslavsky \cite{Zas,Zas2,Zas3}
proved that the chaotic dynamics can be described by using 
the Fokker-Planck equation with the coordinate fractional derivatives.
The fractional Zaslavsky's equation can be used to describe  
dynamics on fractal \cite{Mand,Schr}.
It is known that Fokker-Planck equation can be derived
from the Liouville and BBGKI equations \cite{Is,RL,F}.
The Liouville equation is derived from the
normalization condition and the Hamilton equations \cite{KSF}.
In the Hamilton equations we have only the time derivatives.
Therefore the usual normalization condition leads to the usual
Liouville equation.
The BBGKI equations can be derived from the
Liouville equation and the definition of average value
\cite{Bog3,Gur,Petrina,Born}.
Therefore Zaslavsky's equation \cite{Zas,Zas2,Zas3}
can be derived from a fractional
generalization of the Liouville and BBGKI equations.
To derive the fractional generalization of the Liouville 
and BBGKI equations, 
we use a fractional generalization of normalization condition \cite{chaos} 
and fractional generalization of the definition of average values.
In these generalizations, we use the integrals of fractional order. 
To use these integrals, we must have the physical interpretation 
of the fractional order of integrals. 
We can consider the fractional integrals as 
integrals for the function on a fractional space.
In order to use this interpretation of the fractional integrals,  
we must define a fractional space.
The interpretation of the fractional space
is connected with fractal dimension space.
We can replace the distribution on fractal with fractal mass 
dimension by some continuous distribution that 
is described by fractional integrals.
This procedure is a fractional generalization of Christensen approach \cite{Chr}
that is averaging procedure over fractals.
Suggested procedure leads to the fractional integration and 
differentiation to describe the distributions on fractal.
The fractional integrals allow us to take into account the fractality
of the distribution.
The functions on fractals can be averaged and 
the distribution on fractal can be replaced 
by some "fractional" continuous distribution. 
In order to describe the averaged distribution on fractal  with 
non-integer mass dimension, we must use the fractional integrals.
Smoothing of the microscopic characteristics over the 
physically infinitesimal volume transforms the initial 
distribution on fractal into distribution on space with measure 
that uses the fractional integrals. 
The order of fractional integral is equal 
to the fractal mass dimension.
The consistent approach to describe the distribution on fractal 
is connected with the mathematical definition the integrals
on fractals. In Ref. \cite{RLWQ}, was proved that integrals 
on net of fractals can be approximated by fractional integrals. 
In Ref. \cite{chaos}, we prove that fractional integrals 
can be considered as integrals over the space with fractional 
dimension up to numerical factor. We use the well-known 
formulas of dimensional regularizations \cite{Col}.  
Note that almost all systems with fractional phase space are 
non-Hamiltonian dissipative systems in the usual phase space $(q,p)$.
Therefore we have the other interpretation of the fractional phase 
space. This interpretation follows from the fractional measure
of phase space \cite{chaos} that is used in the fractional integrals.
The fractional phase space can be considered as phase space 
that is described by the fractional powers of coordinates and momenta. 
Using this phase space we can consider
wide class of non-Hamiltonian systems as generalized Hamiltonian systems. 
In this case, the fractional normalization condition 
and the fractional average values are considered as 
condition and values for the generalized Hamiltonian systems 
that are non-Hamiltonian systems in the usual phase space.

In Sec. 2, we consider the fractional generalization of 
normalization condition. 
In Sec. 3, we derive the fractional generalization 
of continuity equation and Liouville equation. 
In Sec. 4,  we consider the physical interpretations of 
fractional phase space  as a space with fractal dimension.
In Sec. 5, the fractional phase space is considered as space with
fractional measure. The fractional generalization 
of the Hamiltonian systems are suggested. 
In Sec. 6, we consider the fractional generalization of average values.
In Sec. 7, the fractional generalization of BBGKI equations are derived 
from the fractional average value and the fractional Liouville equation.
Finally, a short conclusion is given in Sec. 8.

\section{Fractional Generalization of Normalization Condition}

Let us consider a distribution function $\rho(x,t)$ for
$x$ in 1-dimensional Euclidean space $E^1$. 
Let $\rho(x,t) \in L_{1}(E^{1})$, where
$t$ is a parameter. Normalization condition has the form
\[ \int^{+\infty}_{-\infty} \rho(x,t) dx =1. \]
This condition can be rewritten in the form
\be \label{II} \int^{y}_{-\infty} \rho(x,t) dx+
\int^{+\infty}_{y} \rho(x,t) dx=1,\ee
where  $y\in (-\infty,+\infty)$.

Let $\rho(x,t) \in L_{p}(E^{1})$, where $1<p<1/\alpha$.
Fractional integrations on $(-\infty,y)$ and $(y,+\infty)$
are defined \cite{SKM} by the equations
\be \label{I+-} (I^{\alpha}_{+}\rho)(y,t)=\frac{1}{\Gamma(\alpha)}
\int^{y}_{-\infty} \frac{\rho(x,t)dx}{(y-x)^{1-\alpha}} , \quad
(I^{\alpha}_{-}\rho)(y,t)=\frac{1}{\Gamma(\alpha)}
\int^{+\infty}_{y} \frac{\rho(x,t)dx}{(x-y)^{1-\alpha}} . \ee
Using these notations, we rewrite Eq. (\ref{II}) in an equivalent form
\[ (I^{1}_{+}\rho)(y,t)+(I^{1}_{-}\rho)(y,t)=1. \]
The fractional analog of normalization condition (\ref{II}) can be represented
by the following equation  
\[ (I^{\alpha}_{+}\rho)(y,t)+(I^{\alpha}_{-}\rho)(y,t)=1. \]
Equations (\ref{I+-}) can be rewritten in the form
\be \label{Ipm} (I^{\alpha}_{\pm}\rho)(y,t)=
\frac{1}{\Gamma(\alpha)}
\int^{+\infty}_{0} x^{\alpha-1}\rho(y\mp x,t )dx . \ee
This leads to the fractional generalization of  the normalization condition 
\be \label{nc3} \int^{+\infty}_{-\infty} 
\tilde \rho (x,t) d \mu_{\alpha} (x)=1, \ee
where we use the following notations
\be \label{til} \tilde \rho(x,t)=T_x \rho
=\frac{1}{2}\Bigl(\rho(y-x,t)+\rho(y+x,t) \Bigr), \quad
d\mu_{\alpha} (x)=\frac{|x|^{\alpha-1}}{\Gamma(\alpha)} dx.  \ee

The normalization in the phase space is derived by analogy
with a normalization in the configuration space.
The fractional normalization  condition in the phase space is
\be \label{fnc2} \int^{+\infty}_{-\infty} \int^{+\infty}_{-\infty} 
\tilde \rho (q,p,t) d \mu_{\alpha} (q,p)=1, \ee
where $d\mu_{\alpha} (q,p)$ has the form
\be \label{mu0} d\mu_{\alpha} (q,p)=
d\mu_{\alpha} (q) \wedge d\mu_{\alpha} (q)
=\frac{(qp)^{\alpha-1} }{\Gamma^2(\alpha)} dq \wedge dp.  \ee
The distribution function $\tilde \rho(q,p,t)$
in the phase space is defined by
\be \label{trqp} \tilde \rho(q,p,t)=T_q T_p \rho(q,p,t) , \ee
where the operators $T_{q}$ and $T_{p}$ are defined by the equation
\be \label{Txk} T_{x_k} f(...,x_k,...)
=\frac{1}{2}\Bigl( f(...,x^{\prime}_k-x_k,...)
+f(...,x^{\prime}_k+x_k,...) \Bigr) . \ee
This operator allows us to rewrite the distribution functions
\[ \tilde \rho(q,p,t)=
\frac{1}{4} \Bigl(\rho(q'-q,p'-p,t)+\rho(q'+q,p'-p,t)+
\rho(q'-q,p'+p,t)+\rho(q'+q,p'+p,t) \Bigr) \]
in the simple form $\tilde \rho(q,p,t)=T_q T_p \rho(q,p,t)$.

\section{Fractional Generalization of Liouville equation}

\subsection{Fractional Generalization Continuity Equation}

Let us consider a region $W_{0}$ for the time $t=0$.
In the Hamilton picture, we have
\[ \int_{W_{t}} \tilde \rho(x_{t},t) d\mu_{\alpha} (x_{t})=
\int_{W_{0}}\tilde \rho(x_{0},0) d\mu_{\alpha} (x_{0}). \]
Using the replacement of variables $x_{t}=x_{t}(x_{0})$,
where $x_{0}$ is a Lagrangian variable, we get
\[ \int_{W_{0}} \tilde \rho(x_{t},t) |x_t|^{\alpha-1}
\frac{\partial x_{t}}{\partial x_{0}} dx_{0}=
\int_{W_{0}} \tilde \rho(x_{0},0) |x_0|^{\alpha-1} dx_{0}. \]
Since $W_{0}$ is an arbitrary region, we have
\be \label{rmrm} \tilde \rho (x_{t},t) d\mu_{\alpha} (x_{t})=
\tilde \rho (x_{0},0) d\mu_{\alpha} (x_{0}). \ee
This condition leads us to the equation
\[ \tilde \rho(x_{t},t) |x_t|^{\alpha-1}
\frac{\partial x_{t}}{\partial x_{0}} =
\tilde \rho(x_{0},0) |x_0|^{\alpha-1} . \]
Differentiating this equation in time $t$, we obtain
\[ \frac{d \tilde \rho(x_{t},t)}{dt} |x_t|^{\alpha-1}
\frac{\partial x_{t}}{\partial x_{0}}
+ \tilde \rho(x_{t},t) \frac{d}{dt} \Bigl( |x_t|^{\alpha-1}
\frac{\partial x_{t}}{\partial x_{0}}\Bigr)=0 . \]
As the result we have the fractional generalization of the continuity
equation
\be \label{Liu} \frac{d \tilde \rho(x_{t},t)}{dt}
+ \Omega_{\alpha}(x_{t},t) \tilde \rho(x_{t},t)=0 , \ee
where the omega function 
\[ \Omega_{\alpha}(x_{t},t)=
\frac{d}{dt} ln \Bigl( |x_t|^{\alpha-1}
\frac{\partial x_{t}}{\partial x_{0}}\Bigr) \]
describes the velocity of phase volume change.
Eq. (\ref{Liu}) is a fractional continuity equation
in the Hamilton picture.
If the equation of motion has the form
\[ \frac{dx_{t}}{dt}=F_{t}(x), \]
then the function $\Omega_{\alpha}$ is defined by
\be \label{Oax} \Omega_{\alpha}(x_{t},t)=
\frac{d}{dt} \Bigl( ln \ |x|^{\alpha-1}_{t} +
ln \ \frac{\partial x_{t}}{\partial x_{0}}\Bigr)=
\frac{\alpha-1}{x_{t}} \frac{dx_{t}}{dt} +
\frac{\partial}{\partial x_{t}} \frac{dx_{t}}{dt} 
= \frac{(\alpha-1)F_{t}}{x_{t}} +
\frac{\partial F_{t}}{\partial x_{t}}. \ee

\subsection{Fractional Continuity Equation for Phase Space}

Using phase space analog of Eq. (\ref{rmrm}) in the form 
\[ \tilde \rho_t d\mu_{\alpha} (q_t,p_t)
=\tilde \rho_0 d\mu_{\alpha} (q_0,p_0) , \]
we get the relation
\be \label{mu3}
\tilde \rho_t \frac{|q_tp_t|^{\alpha-1}}{ \Gamma^2(\alpha)}
dq_t \wedge d p_t =\tilde \rho_0
\frac{|q_0p_0|^{\alpha-1}}{ \Gamma^2(\alpha)}
dq_0 \wedge d p_0  . \ee
Let us use the well-known transformation
\be \label{qpt0} dq_t \wedge d p_t
= \{q_t,p_t\}_0 dq_0 \wedge d p_0 ,\ee
where $\{q_t,p_t\}_0$ is Jacobian which
is defined by the determinant
$$\{q_t,p_t\}_0 =det  \frac{\partial(q_t,p_t)}{\partial (q_0,p_0)}
=det \ \left( \begin{array}{cc}
{\partial q_{kt}}/{\partial q_{l0}}&
{\partial q_{kt}}/{\partial p_{l0}}\\
{\partial p_{kt}}/{\partial q_{l0}}&
{\partial p_{kt}}/{\partial p_{l0}}
\end{array}
\right).$$
As the result, we have condition (\ref{mu3}) in the form
\be \label{nn1} \tilde \rho_t |q_tp_t|^{\alpha-1}\{q_t,p_t\}_0 =
|q_0p_0|^{\alpha-1} \tilde \rho_0 . \ee
This equation can be rewritten in more simple form
\be \label{nn2} \tilde \rho_t \{q^{\alpha}_t,p^{\alpha}_t\}_0 =
\alpha^2 |q_0p_0|^{\alpha-1} \tilde \rho_0 . \ee
We use the following notation for fractional power of
coordinates and momenta
\be \label{xa}
x^{\alpha} =\beta(x) (x)^{\alpha}= sgn(x) |x|^{\alpha} , \ee
where $\beta(x)=(sgn(x))^{\alpha-1}$. 
The function $sgn(x)$ is equal to $+1$ for $x\ge0$, and $-1$ for $x<0$.
The total time derivatives of Eq. (\ref{nn2}) lead us to the fractional 
generalization of Liouville equation in the form
\be \label{Liu3} \frac{d \tilde \rho }{dt}
+\Omega_{\alpha} \tilde \rho =0 , \ee
where the omega function $\Omega_{\alpha}$ is defined by
\be \label{02}
\Omega_{\alpha}= \{q^{\alpha}_t,p^{\alpha}_t\}^{-1}_0
\frac{d}{dt}\{q^{\alpha}_t,p^{\alpha}_t\}_0=
\frac{d}{dt} ln \{q^{\alpha}_t,p^{\alpha}_t\}_0 . \ee
In the usual notations, we have
\be \label{o3} \Omega_{\alpha}=\frac{d}{dt} ln \ det
\frac{\partial(q^{\alpha}_t,p^{\alpha}_t)}{\partial (q_0,p_0)} . \ee
Using well-known relation $ln\ det \ A=Sp \ ln \ A$,
we can write the omega function $\Omega_{\alpha}$ in the form
\[ \Omega_{\alpha}=\{\frac{dq^{\alpha}_t}{dt},p^{\alpha}_t\}_{\alpha}+
\{q^{\alpha}_t,\frac{dp^{\alpha}_t}{dt}\}_{\alpha} , \]
where $\{ \ ,\ \}_{\alpha}$ is the fractional generalization 
of the Poisson brackets in the form
\[ \{A,B\}_{\alpha}=\frac{\partial A}{\partial q^{\alpha}}
\frac{\partial B}{\partial p^{\alpha}}-
\frac{\partial A}{\partial p^{\alpha}}
\frac{\partial B}{\partial q^{\alpha}}  . \]
In the general case ($\alpha \not=1$), the function
$\Omega_{\alpha}$ is not equal to zero ($\Omega_{\alpha} \not=0$)
for the systems that are Hamiltonian systems in the usual phase space.
If $\alpha=1$, we have $\Omega_{\alpha} \not=0$ only
for non-Hamiltonian systems.
If the Hamilton equations have the form
\be \label{H3}
\frac{dq_t}{dt}=K(q_t,p_t), \quad \frac{dp_t}{dt}=F(q_t,p_t), \ee
then the omega function $\Omega_{\alpha}$ is defined by
\be \label{om3}  \Omega_{\alpha}(q,p)=
(\alpha-1)\Bigl(q^{-1} K(q,p)+p^{-1} F(q,p)\Bigr)+
\{K,p\}_1+\{q,F\}_1. \ee
This relation allows to derive $\Omega_{\alpha}$
for all dynamical systems (\ref{H3}).
It is easy to see that the usual nondissipative system
\be \label{eq1} \frac{dq}{dt}=\frac{p}{m},
\quad \frac{dp}{dt}=f(q) \ee
has the omega function
\[ \Omega_{\alpha}(q,p)=(\alpha-1)(mqp)^{-1}(p^2+mq f(q)) , \]
and can be considered as a dissipative system in the fractional 
phase space $(q^{\alpha},p^{\alpha})$.

\subsection{Fractional Liouville Equation for N-particle System}

Let us consider $N$-particle system.
Suppose $k$-particle is described by the
generalized coordinates ${\bf q}_{k}=(q_{k1},...,q_{kn})$ and 
generalized momenta ${\bf p}_{k}=(p_{k1},...,p_{kn})$ 
that satisfy the Hamilton equations in the form
\be \label{H3b}
\frac{d{\bf q}^{\alpha}_{k}}{dt}=
{\bf K}_k({\bf q}^{\alpha},{\bf p}^{\alpha}), \quad
\frac{d{\bf p}^{\alpha}_{k}}{dt}=
{\bf F}_k({\bf q}^{\alpha},{\bf p}^{\alpha},t). \ee
Here, we use the notation (\ref{xa}) for fractional power of
coordinates and momenta. 
The evolution of N-particle distribution function $\rho_{N}$
is described by the Liouville equation.
Using the fractional normalization condition 
\be \hat I^{\alpha}[1,...,N] \tilde \rho_{N}({\bf q},{\bf p},t)=1 , \ee
we can derive the fractional Liouville equation \cite{chaos} 
for N-particle distribution function in the form
\be \label{L1}
\frac{d \tilde \rho_{N}}{dt}+ \Omega_{\alpha} \tilde \rho_{N} =0, \ee
where $d/dt$ is a total time derivative
\[ \frac{d}{dt}=\frac{\partial}{\partial t}+
\sum^{N}_{k=1}\frac{d{\bf q}_{k}}{dt}\frac{\partial}{\partial {\bf q}_{k}}+
\sum^{N}_{k=1}\frac{d{\bf p}_{k}}{dt}\frac{\partial}{\partial {\bf p}_{k}} . \]
Here, we use the following notations
\[ {\bf A}_k {\bf B}_k=\sum^{n}_{a=1} A_{ka} B_{ka} . \]
Using Eq. (\ref{H3b}), the total time derivative can be written 
for the fractional powers in the form
\be \label{ttd3} \frac{d}{dt}=\frac{\partial}{\partial t}+
\sum^{N}_{k=1}
{\bf K}_k\frac{\partial}{\partial {\bf q}^{\alpha}_{k}}+
\sum^{N}_{k=1}
{\bf F}_k\frac{\partial}{\partial {\bf p}^{\alpha}_{k}} . \ee
The omega function $\Omega_{\alpha}$ is defined by the equation
\be  \label{o2} \Omega_{\alpha}=\sum^{N}_{k=1}\sum^{n}_{a=1}\Bigl(
\{ K^k_a,p^{\alpha}_{ka}\}_{\alpha}+
\{q^{\alpha}_{ka},F^k_a\}_{\alpha} \Bigr) . \ee
Here, we use the following notations for the brackets
\be \label{PB0} \{A,B\}_{\alpha}=\sum^{N}_{k=1}\sum^{n}_{a=1}\Bigl(
\frac{\partial A}{\partial q^{\alpha}_{ka}}
\frac{\partial B}{\partial p^{\alpha}_{ka}}-
\frac{\partial A}{\partial p^{\alpha}_{ka}}
\frac{\partial B}{\partial q^{\alpha}_{ka}} \Bigr) . \ee
Using Eqs. (\ref{L1}), (\ref{o2}) and (\ref{ttd3}), we can rewrite
the Liouville equation in the equivalent form
\be \label{r2}
\frac{\partial \tilde \rho_{N}}{\partial t}={\cal L}_{N} \tilde \rho_{N} , \ee
where ${\cal L}_{N}$ is Liouville operator that is
defined by the equation
\be \label{Lam} {\cal L}_{N} \tilde \rho_{N} =-
\sum^{N}_{k=1} \Bigl(
\frac{\partial ({\bf K}_k \tilde \rho_{N})}{\partial {\bf q}^{\alpha}_{k}}+
\frac{\partial ({\bf F}_k \tilde \rho_{N})}{\partial {\bf p}^{\alpha}_{k}}
 \Bigr) . \ee

\section{Fractional Space as Space with Fractal Dimension}

The interpretation of the fractional space
can be connected with fractal mass dimension.
Fractal dimension can be best calculated by box counting 
method which means drawing a box of size $R$ 
and counting the mass inside. 
The mass fractal dimension \cite{Mand,Schr} can be easy measured.
The properties of the fractal like mass obeys a power law relation
\be \label{MR} M(R) =kR^{{\alpha}} ,   \ee
where $M$ is the mass of fractal, $R$ is a box size (or a sphere radius),
and ${\alpha}$ is a mass fractal dimension. 
Amount of mass inside a box of size $R$
has a power law relation (\ref{MR}).

Let us consider the region $W$ in 3-dimensional Euclidean space $E^3$.
The volume of the region $W$ is denoted by $V(W)$.
The mass of the region $W$ of the distribution on fractal 
is denoted by $M(W)$. 
The fractality means that the mass of this distribution  
in any region $W$ of Euclidean space $E^3$ increase more slowly 
that the volume of this region $E^3$.
For the ball region of the distribution on fractal, 
this property can be described by the power law (\ref{MR}), 
where $R$ is the radius of the ball $W$. 

The distribution on fractal is called homogeneous if the power law (\ref{MR}) 
does not depends on the translation and rotation of the region $W$. 
The homogeneity property of the distribution can be formulated in the form:
For all regions $W_1$ and $W_2$ such that 
the volumes are equal $V(W_1)=V(W_2)$, 
we have that the mass of these regions are equal $M(W_1)=M(W_2)$. 
In order to describe the homogeneous distribution on fractals, we must use the 
continuous distribution such that fractality 
and homogeneity properties can be realized in the form: \\

\noindent
(1) Fractality:
The mass of the ball region $W$ for the distribution on fractal 
obeys a power law relation (\ref{MR}). 
In the general case, we have the scaling law relation
\[ dM_{\alpha}(\lambda W)=\lambda^{\alpha} dM_{\alpha}(W) , \]
where $\lambda W=\{\lambda x, \ \ x \in W \}$. \\

\noindent
(2) Homogeneity:
The local density of homogeneous fractal distribution
is translation and rotation invariant value that have the form
$\rho(x)=\rho_0=const$.

\vskip 3mm

We can realize these requirements by the 
fractional generalization of the equation
\be \label{MW} M_3(W)=\int_W \rho(x) d^3 x . \ee
Let us define the fractional integral
in Euclidean space $E^3$ in the Riesz form
\cite{SKM} by the equation
\be \label{ID} (I^{{\alpha}}\rho)(y)=
\int_W \rho(x) d\mu_{\alpha}(x) , \ee
where $d\mu_{\alpha}(x)=c_3({\alpha},x,y)d^3 x$, and 
\[ c_3({\alpha},x,y)=\frac{2^{3-{\alpha}} \Gamma(3/2)}{\Gamma({\alpha}/2)} 
|x-y|^{{\alpha}-3} ,  \quad
|x-y|=\sqrt{\sum^3_{k=1} (x_k-y_{k0})^2}. \]
The point $y \in W$ is the initial point of the fractional integral.
We will use the initial points in the integrals are set to zero ($y=0$).
The numerical factor in Eq. (\ref{ID}) has this form in order to
derive usual integral in the limit ${\alpha}\rightarrow (3-0)$.
Note that the usual numerical factor
$\gamma^{-1}_3({\alpha})=
{\Gamma(1/2)}/{2^{\alpha} \pi^{3/2} \Gamma({\alpha}/2)}$,
which is used in Ref. \cite{SKM},  
leads us to $\gamma^{-1}_3(3-0)= {\Gamma(1/2)}/{2^3 \pi^{3/2} \Gamma(3/2)}$ 
in the limit ${\alpha}\rightarrow (3-0)$. 

Using notations (\ref{ID}), we can rewrite Eq. (\ref{MW})
in the form $M_3(W)=(I^{3}\rho)(y)$. 
Therefore the fractional generalization of this equation can be
defined in the form
\be \label{MWD}  M_{\alpha}(W)=(I^{\alpha} \rho)(y)=
\frac{2^{3-{\alpha}} \Gamma(3/2)}{\Gamma({\alpha}/2)}
\int_W \rho(x) |x-y|^{{\alpha}-3} d^3 x . \ee

If we consider the homogeneous fractal distribution 
($\rho(x)=\rho_0=const$) and the ball region $W=\{x: \  |x|\le R \}$, then
we have 
\[ M_{\alpha}(W)= \rho_0 \frac{2^{3-{\alpha}} \Gamma(3/2)}{\Gamma({\alpha}/2)} 
\int_W |x|^{{\alpha}-3} d^3 x . \]
Using the spherical coordinates, we get
\[ M_{\alpha}(W)= \frac{\pi 2^{5-{\alpha}} \Gamma(3/2)}{\Gamma({\alpha}/2)} 
\rho_0 \int_W |x|^{{\alpha}-1} d |x|= 
\frac{2^{5-{\alpha}} \pi \Gamma(3/2)}{{\alpha} \Gamma({\alpha}/2)} 
\rho_0 R^{{\alpha}} . \]
As the result, we have $M(W)\sim R^{\alpha}$, i.e., we derive Eq. (\ref{MR})
up to the numerical factor.
Therefore the distribution on fractal 
with non-integer mass dimension ${\alpha}$ 
can be described by fractional integral of order ${\alpha}$.

Note that the interpretation of the fractional integration
is connected with fractional dimension \cite{chaos}.
This interpretation follows from
the well-known formulas for dimensional regularizations \cite{Col}:
\be \label{dr} \int f(x) d^{{\alpha}} x =
\frac{2 \pi^{{\alpha}/2}}{\Gamma({\alpha}/2)}
\int^{+\infty}_{0} f(x)  x^{{\alpha}-1} dx  . \ee
Using Eq. (\ref{dr}), we get
that the fractional integral 
\[ \int_W f(x) d\mu_{\alpha}(x), \]
can be considered as integral in the fractional dimension space
\be \label{fnc-2} 
\frac{\Gamma({\alpha}/2) }{2 \pi^{{\alpha}/2} \Gamma({\alpha})}
\int  f(x) d^{{\alpha}} x  \ee
up to the numerical factor
$\Gamma({\alpha}/2) /( 2 \pi^{{\alpha}/2} \Gamma({\alpha}))$.

\section{Fractional Space as Space with Fractional Measure}

The interpretation of the fractional space is
connected with the fractional measure 
that is used in the fractional integrals. 
The parameter $\alpha$  defines the space with
the fractional measure (volume) of the region $W$.
It is easy to prove that the velocity of the fractional 
measure (volume) change is defined by the omega function (\ref{Oax}).

\subsection{Fractional Phase Volume for Configuration Space}

The usual phase volume of the region $W=\{x:\ x\in [a;b]\}$ 
in Euclidean space $E^1$ is defined by
\be \label{V1} \mu_1(W) =\int^b_a dx =\int^y_a dx +\int^b_y dx  , \ee
where $y \in [a;b]$. 
Using the fractional integrals \cite{SKM} in the form 
\[ (I^{\alpha}_{a+}1)(y)=
\frac{1}{\Gamma (\alpha)} \int^{y}_{a} \frac{dx}{(y-x)^{1-\alpha}}, \quad
(I^{\alpha}_{b-}1)(y)=
\frac{1}{\Gamma (\alpha)} \int^{b}_{y} \frac{dx}{(x-y)^{1-\alpha}}, \]
we get the phase volume (\ref{V1}) in the equivalent form
\be \label{V2} \mu_1(W) =(I^{1}_{a+}1)(y)+(I^{1}_{b-}1)(y) . \ee
The fractional generalization of the phase volume can be defined by
\be \label{Valpha} \mu_{\alpha}(W)=(I^{\alpha}_{a+}1)(y)
+(I^{\alpha}_{b-}1)(y) . \ee
The fractional phase volume integral can be represented 
by the equation 
\be \label{PV4} \mu_{\alpha}(W)=
\int^{+(b-y)}_{-(b-y)}g(\alpha) d \mu_{\alpha}(x) . \ee
Here, we use the notations 
\be d \mu_{\alpha}(x)=\frac{|x|^{\alpha-1}dx}{\Gamma(\alpha)}=
\frac{dx^{\alpha}}{\alpha \Gamma(\alpha)}, \qquad 
g(\alpha)=\frac{1}{2} \Bigl(1+\Bigl(\frac{y-a}{b-y}\Bigr)^{\alpha})
, \ee
We use the notation (\ref{xa}) for fractional power of
coordinates.

\subsection{Fractional Phase Volume for Phase Space}

The fractional measure for the region $W$ of 2n-dimensional phase 
space can be defined by the equation
\be  \label{sim1} \mu_{\alpha}(W)=
\int_W g(\alpha) d\mu_{\alpha}(q,p), \ee
where $d\mu_{\alpha}(q,p)$ is a phase volume element
\be \label{muV} d\mu_{\alpha}(q,p)=
\prod^n_{k=1} \frac{dq^{\alpha}_k \wedge d p^{\alpha}_k}{
(\alpha \Gamma(\alpha))^{2}} , \ee
and $g(\alpha)$ is a numerical multiplier.
If the region $W$ of the phase space is defined by
$q_k\in E^1$ and $p_k\in E^1$, then $g(\alpha)=1/4^n$.
If this region is defined by $q_k\in [q_{ak};q_{bk}]$
and $p_k\in [p_{ak};p_{bk}]$, then
\be g(\alpha)=\frac{1}{4^{n}} \prod^{n}_{k=1} g_k(\alpha),  \quad
g_k(\alpha)=
\Bigl(1+\Bigl(\frac{q_{bk}-y_k}{y_k-q_{ak}}\Bigr)^{\alpha}\Bigr)
\Bigl(1+\Bigl(\frac{p_{bk}-y'_k}{y'_k-p_{ak}}\Bigr)^{\alpha}\Bigr) \ee
It is easy to see that the fractional measure depends on the fractional
powers of coordinates and momenta $(q^{\alpha},p^{\alpha})$. 

For example,  the phase volume for
the two-dimensional phase space has the form
\be \label{PV5} \mu_{\alpha}(W)= \int^{(q_b-y)}_{-(q_b-y')}
\int^{(p_b-y)}_{-(p_b-y')} g_1(\alpha)
\frac{dq^{\alpha} \wedge d p^{\alpha}}{(\alpha \Gamma(\alpha))^{2}} 
=\int^{(q_b-y)}_{-(q_b-y')}
\int^{(p_b-y)}_{-(p_b-y')} g_1(\alpha) |q p|^{\alpha-1} 
\frac{dq \wedge d p}{(\Gamma(\alpha))^{2}} . \ee

Note that the volume element of fractional phase space can be realized 
by fractional exterior derivatives \cite{CN} that are defined by
\[ d^{\alpha}=\sum^{n}_{k=1}dq^{\alpha}_k \frac{\partial^{\alpha}}{(\partial 
(q_k-y_k))^{\alpha}}+\sum^{n}_{k=1}
dp^{\alpha}_k \frac{\partial^{\alpha}}{(\partial (p_k-{y'}_k))^{\alpha}}. \]
For example, the two-dimensional phase space is defined by
\[ d \mu_{\alpha}(q,p)=\frac{1}{(\alpha \Gamma(\alpha))^2}
dq^{\alpha} \wedge dp^{\alpha}= 
\frac{1}{(\alpha \Gamma(\alpha))^2}
\Bigl(\frac{4}{\Gamma^2(2-\alpha)}+
\frac{1}{\Gamma^2(1-\alpha)}\Bigr)^{-1} 
(qp)^{\alpha-1} d^{\alpha}q \wedge d^{\alpha}p . \]

\subsection{Fractional Phase Volume Change}

The interpretation of the fractional phase space
is connected with the fractional measure of phase space.
The parameter $\alpha$  defines the space with
the fractional phase measure (\ref{sim1}) and (\ref{muV}). 
It is easy to prove that the velocity of the fractional 
phase volume change is defined by the equation
\[ \frac{d\mu_{\alpha}(W)}{dt}=
\int_W \Omega_{\alpha}(q,p,t) g(\alpha) d\mu_{\alpha}(q,p) , \]
where the omega function $\Omega_{\alpha}$ has the form 
\be \label{Oa} \Omega_{\alpha}=
\{\frac{dq^{\alpha}_t}{dt},p^{\alpha}_t\}_{\alpha}+
\{q^{\alpha}_t,\frac{dp^{\alpha}_t}{dt}\}_{\alpha} . \ee
Here, we use the fractional brackets:
\be \label{PB} \{A,B\}_{\alpha}=
\sum^n_{k=1}\Bigl(\frac{\partial A}{\partial q^{\alpha}_{k}}
\frac{\partial B}{\partial p^{\alpha}_{k}}-
\frac{\partial A}{\partial p^{\alpha}_{k}}
\frac{\partial B}{\partial q^{\alpha}_{k}} \Bigr) , \ee
where we use notations (\ref{xa}).

The form of the omega function allows us to consider 
new class of the systems that are described by 
the fractional powers of coordinates and momenta.
The system can be called a fractional dissipative system
if a fractional phase volume changes, i.e., $\Omega_{\alpha} \not=0$.
The system which is a nondissipative system
in the usual phase space,
can be a dissipative system in the fractional phase space.
The usual nondissipative systems (\ref{eq1})
are dissipative in the fractional phase space.

\subsection{Fractional Generalization of Hamiltonian Systems}

Fractional phase space can be considered as a phase space of the 
systems that are described by the fractional powers of coordinates and momenta.
Let us consider the class of non-Hamiltonian systems that are described 
by the fractional powers of coordinates and momenta.
A system is called a fractional system if the phase space of the system
is described by the fractional powers of coordinates  
and momenta 
\be \label{qapa} q^{\alpha}_k =\beta(q) (q_k)^{\alpha}= 
sgn(q_k) |q_k|^{\alpha}, \quad
p^{\alpha}_k =\beta(p) (p_k)^{\alpha}= sgn(p_k) |p_k|^{\alpha}. \ee
Here $k=1,...,n$, and $\beta(x)$ is defined by Eq. (\ref{xa}).
We can consider the fractional systems in the usual phase space $(q,p)$
and in the fractional phase space $(q^{\alpha},p^{\alpha})$.
In the second case, the equations of motion for the fractional 
systems have more simple form. 
Therefore we use the fractional phase space that is a space with the 
fractional measure  that is used in the fractional integrals. 
We consider the fractional power of the coordinates 
as a convenient way to describe systems in the space with measure
that is defined by fractional integrals. 

A classical system (in the usual phase space) 
is called Hamiltonian if the right-hand sides of the equations
\be \label{qg-pf} \frac{dq_{k}}{dt}=K_k(q,p),
\quad \frac{dp_{k}}{dt}=F_k(q,p) \ee
satisfy the following Helmholtz conditions \cite{Tar-tmf3}:
\be \label{HelmCond}
\frac{\partial K_k}{\partial p_l}-\frac{\partial K_l}{\partial p_k}=0,
\quad
\frac{\partial K_k}{\partial q_l}-\frac{\partial F_l}{\partial p_k}=0,
\quad
\frac{\partial F_k}{\partial q_l}-\frac{\partial F_l}{\partial q_k}=0.
\ee
In this case, we can rewrite Eqs. (\ref{qg-pf}) in the form
\[ \frac{dq_{k}}{dt}=\frac{\partial H}{\partial p_{k}},
\quad \frac{dp_{k}}{dt}=-\frac{\partial H}{\partial q_{k}} . \]
Using the Poisson brackets $\{\ , \ \}_1$, 
we can rewrite these equations in an equivalent form
\[ \frac{dq_{k}}{dt}=\{q_k,H\}_1 , \quad \frac{dp_{k}}{dt}=\{p_k,H\}_1 . \]

The fractional generalization of the Hamiltonian system 
is described by the equation
\be \label{33} \frac{dq^{\alpha}_{k}}{dt}=
\frac{\partial H}{\partial p^{\alpha}_{k}},
\quad \frac{dp^{\alpha}_{k}}{dt}=
-\frac{\partial H}{\partial q^{\alpha}_{k}}, \ee
where $H$ is a function that can be considered as a 
fractional analog of the Hamiltonian.
Note that the function $H$ such that $\partial H / \partial t=0$ 
is the invariant of the motion. 
Using the brackets (\ref{PB}), we can rewrite Eq. (\ref{33}) in
the equivalent form 
\be \frac{dq^{\alpha}_{k}}{dt}=\{q^{\alpha}_{k},H\}_{\alpha},
\quad \frac{dp^{\alpha}_{k}}{dt}=\{p^{\alpha}_{k},H\}_{\alpha}. \ee
These equations describe the system in the fractional phase space 
$(q^{\alpha},p^{\alpha})$.  
For the usual phase space $(q,p)$, the fractional Hamiltonian systems
(\ref{33}) are described by the equations
\be \label{35} \frac{dq_{k}}{dt}=\frac{(q_kp_k)^{1-\alpha}}{\alpha^2}
\frac{\partial H}{\partial p_{k}},
\quad \frac{dp_{k}}{dt}=- \frac{(q_kp_k)^{1-\alpha}}{\alpha^2}
\frac{\partial H}{\partial q_{k}}. \ee
The fractional Hamiltonian systems (\ref{35}) are non-Hamiltonian
systems in the usual phase space $(q,p)$. 
It is easy to prove that the Helmholtz conditions (\ref{HelmCond})
are not satisfied. 
Therefore fractional Hamiltonian system (\ref{35})
is a non-Hamiltonian system in the usual phase space $(q,p)$.
The fractional phase space allows us to write the equations 
of motion for the non-Hamiltonian systems (\ref{35})
in the generalized Hamiltonian form (\ref{33}).

The omega function for the system (\ref{qg-pf}) 
in the usual phase space $(q,p)$ is defined by the equation
\be \label{51}
\Omega=\sum^n_{k=1}\Bigl( \frac{\partial K_k}{\partial q_k}+
\frac{\partial F_k}{\partial p_k}\Bigr) .
\ee
If the omega function is negative $\Omega<0$, then 
the system is called a dissipative system.
If $\Omega \not=0$, then the system is a generalized
dissipative system.
For the fractional Hamiltonian systems (\ref{35}), 
the omega function (\ref{51}) is not equal to zero.
Therefore the fractional Hamiltonian systems are the 
general dissipative systems in the usual phase space.

The function $H=H(q^{\alpha},p^{\alpha})$ can be considered as 
a fractional analog of the Hamiltonian function. For example, 
we can use 
\be \label{Ha} H(q^{\alpha},p^{\alpha})=
\sum^n_{k=1} \frac{p^{2\alpha}_{k}}{2m} +U(q^{\alpha}) . \ee
It is easy to see that fractional systems with Hamiltonian (\ref{Ha})
lead us to the non-Gaussian statistics.
The interest in and relevance of fractional kinetic equations
is a natural consequence of the realization of the importance of
non-Gaussian statistics of many dynamical systems. There is already
a substantial literature studying such equations in one or more
space dimensions.

Note that the classical dissipative non-Hamiltonian systems can have
canonical Gibbs distribution as a solution of stationary
Liouville equations for this dissipative system \cite{Tar-mplb}.
Using the methods \cite{Tar-mplb}, it is easy to prove
that some of fractional dissipative systems can have
fractional generalization of the canonical Gibbs distribution 
in the form
\[ \rho(q,p)=Z(T)exp -\frac{H(q^{\alpha},p^{\alpha})}{kT} , \]
as a solution of the fractional Liouville equations
\be \label{LE} \frac{\partial \rho}{\partial t}+
\frac{p^{\alpha}_{k}}{m}\frac{\partial \rho}{\partial q^{\alpha}_{k}}+
\frac{\partial}{\partial p^{\alpha}_{k}}
\Bigl(F_{k}(q^{\alpha},p^{\alpha}) \rho\Bigr)=0 . \ee
Here the function $H(q^{\alpha},p^{\alpha})$ is defined by (\ref{Ha}).

\section{Fractional Generalization of Average Values}

The usual average value  for the configuration space
\be <f>_1= \int^{+\infty}_{-\infty} f(x,t)\rho(x,t) dx  \ee
can be written in the form
\be \label{If}
<f>_1=\int^y_{-\infty} f(x,t)\rho(x,t) dx +
\int^{+\infty}_y f(x,t)\rho(x,t) dx , \ee
where $y\in (-\infty,+\infty)$. Using the notations (\ref{I+-}),
we can rewrite the average value (\ref{If}) in the form
\[ <f>_1(y,t)=(I^{1}_{+}f\rho)(y,t)+(I^{1}_{-}f\rho)(y,t) . \]
The fractional generalization of
this equation is defined by
\be \label{Aa} <f>_{\alpha}(y,t)=
(I^{\alpha}_{+}f\rho)(y,t)+(I^{\alpha}_{-}f\rho)(y,t) . \ee
We can rewrite Eq. (\ref{Aa}) in the form
\be \label{FI5} <f>_{\alpha}(y,t)= 
\int^{+\infty}_{-\infty} T_x (f\rho)(x,t) d\mu_{\alpha}(x) , \ee
where we use
\be \label{dm}
d\mu_{\alpha}(x)=\frac{|x|^{\alpha-1} dx}{\Gamma(\alpha)}=
\frac{d x^{\alpha}}{\alpha \Gamma(\alpha)} ,\quad
T_x f=\frac{1}{2}\Bigl(f(y-x)+f(y+x)\Bigr) , \ee
and $x^{\alpha}$ is defined by Eq. (\ref{xa}).

We can define the integral operator $\hat I^{\alpha}_{x}$ by the equation
\be \hat I^{\alpha}_{x} f(x)=
\int^{+\infty}_{-\infty}  f(x) d \mu_{\alpha} (x) .\ee
In this case, the fractional generalization of average value (\ref{FI5})  
can be written in the form
\[ <f>_{\alpha}(y,t)=\hat I^{\alpha}_{x} T_x f(x)\rho(x,t) . \]

Let us consider $k$-particle that is described by
generalized coordinates ${\bf q}_{k}=(q_{k1},...,q_{kn})$ and generalized
momenta ${\bf p}_{k}=(p_{k1},...,p_{kn})$.
We can define the phase space integral operator for $k$-particle by the equation
\[ \hat I^{\alpha}[k]=
\hat I^{\alpha}_{q_{k1}} \hat I^{\alpha}_{p_{k1}} ...
\hat I^{\alpha}_{q_{kn}} \hat I^{\alpha}_{p_{kn}} . \] 
For the $N$-particle system, we use the operators
\[ \hat I^{\alpha}[1,...,N]=\hat I^{\alpha}[1]...\hat I^{\alpha}[N] ,\quad
T[1,...,N]=T[1]...T[N] , \] 
where the operator $T[k]$ is defined by the relation
\[ T[k]=T_{q_{k1}} T_{p_{k1}}...T_{q_{kn}} T_{p_{kn}} . \]
Here the operator $T_{xk}$ is defined by Eq. (\ref{dm}). 

Using these notations, we can define
fractional analog of the average values 
for the phase space of $N$-particle system by the relation
\be  \label{fa} <f>_{\alpha}(y,t)=
\hat I^{\alpha}[1,...,N] T[1,...,N] f \rho_{N} .  \ee
In the simple case ($N=m=1$), the fractional average 
value (\ref{fa}) is defined by the equation 
\be \label{AV2} <f>_{\alpha}(y,t)=
\int^{+\infty}_{-\infty} \int^{+\infty}_{-\infty}
d\mu_{\alpha}(q,p) \ T_q T_p f(q,p)\rho(q,p,t) . \ee
Note that the fractional normalization condition 
is a special case of this definition of the average value
$<1>_{\alpha}(y,t)=1$.

\section{Fractional Generalization of BBGKI Equations}

The state of N-particle system is described by 
N-particle distribution function 
\[ \rho_{N}({\bf q},{\bf p},t)=
\rho({\bf q}_{1},{\bf p}_{1},...,{\bf q}_{N},{\bf p}_{N},t) .  \]
We use the tilde distribution functions
\be \tilde \rho_{N}({\bf q},{\bf p},t)=
T[1,...,N]\rho_{N}({\bf q},{\bf p},t) . \ee
The fractional generalization of 1-particle reduced distribution function
$\tilde \rho_1$ can be defined by the equation 
\be \label{r1} \tilde \rho_{1}({\bf q},{\bf p},t)=
\tilde \rho({\bf q}_{1},{\bf p}_{1},t)=
\hat I^{\alpha}[2,...,N]\tilde \rho_{N}({\bf q},{\bf p},t). \ee
Obviously, that 1-particle distribution function satisfies
the fractional normalization condition 
\be \label{r3} 
\hat I^{\alpha}[1] \tilde \rho_{1}({\bf q},{\bf p},t)=1 . \ee

The BBGKI equations \cite{Bog3,Gur,Petrina,Born}
are equations for the reduced distribution functions.
These equations can be derived from the Liouville equation.
Let us derive the fractional generalization of the first BBGKI equation 
from the fractional Liouville equation (\ref{r2}).

In order to derive the equation for 1-particle distribution 
function $\tilde \rho_{1}$ we differentiate Eq. (\ref{r1}) 
with respect to time: 
\[ \frac{\partial \tilde \rho_{1}}{\partial t}=
 \frac{\partial}{\partial t} \hat I^{\alpha}[2,...,N] \tilde \rho_{N}=
\hat I^{\alpha}[2,...,N] \frac{\partial \tilde \rho_{N}}{\partial t} . \]
Using the Liouville equation (\ref{r2}) for
$N$-particle distribution  function $\tilde \rho_{N}$, we have
\be \label{92a} \frac{\partial \tilde \rho_{1}}{\partial t}=
\hat I^{\alpha}[2,...,N] {\cal L}_{N} \tilde \rho_{N}({\bf q},{\bf p},t) . \ee
Substituting (\ref{Lam}) in Eq. (\ref{92a}), we get
\be \label{r1i}
\frac{\partial \tilde \rho_{1}}{\partial t}=- \hat I^{\alpha}[2,...,N]
\sum^{N}_{k=1} \Bigl(
\frac{\partial ({\bf K}_{k} \tilde \rho_{N})}{\partial {\bf q}^{\alpha}_{k}}
+\frac{\partial ({\bf F}_{k}\tilde \rho_{N})}{\partial {\bf p}^{\alpha}_{k}}
\Bigr) . \ee

Let us consider in Eq. (\ref{r1i}) the integration over
${\bf q}_{k}$ and ${\bf p}_{k}$ for k-particle term. 
Since the coordinates and momenta are independent variables, 
we can derive
\be \hat I^{\alpha}[{\bf q}_k]
\frac{\partial }{\partial {\bf q}^{\alpha}_{k}} ({\bf K}_k \tilde \rho_{N}) =
\frac{1}{\alpha \Gamma(\alpha)}
\Bigl({\bf K}_k \tilde \rho_{N} \Bigr)^{+\infty}_{-\infty}=0 . \ee
Here, we use that the distribution $\tilde \rho_{N}$ in the limit
${\bf q}_k \rightarrow \pm \infty$ is equal to zero.
It follows from the normalization condition.  
If the limit is  not equal to zero, then
the integration over phase space is equal to infinity.
Similarly, we have
\[ \hat I^{\alpha}[{\bf p}_{k}] 
\Bigl( \frac{\partial}{\partial {\bf p}^{\alpha}_{k}}
({\bf F}_{k} \tilde \rho_{N}) \Bigr) =
\frac{1}{\alpha \Gamma(\alpha)}
\Bigl({\bf F}_{k} \tilde \rho_{N} \Bigr)^{+\infty}_{-\infty}=0  . \]
Then all terms in Eq. (\ref{r1i}) with $k=2,...,n$
are equal to zero. We have only term for $k=1$.
Therefore Eq. (\ref{r1i}) has the form
\be \label{r1i2} \frac{\partial \tilde \rho_{1}}{\partial t}=
- \hat I^{\alpha}[2,...,N]\Bigl(
\frac{\partial ({\bf K}_1 \tilde \rho_{N})}{\partial {\bf q}^{\alpha}_{1}}
+\frac{\partial ({\bf F}_{1} \tilde \rho_{N})}{\partial {\bf p}^{\alpha}_{1}}
\Bigr) . \ee

Since the variable ${\bf q}_{1}$ is an independent of
${\bf q}_{2},...,{\bf q}_{N}$ and  ${\bf p}_{2},...,{\bf p}_{N}$,
the first term in Eq. (\ref{r1i2}) can be written in the form
\[ \hat I^{\alpha}[2,...,N]
\frac{\partial ({\bf K}_1 \tilde \rho_{N})}{\partial {\bf q}^{\alpha}_{1}} 
= \frac{\partial}{\partial {\bf q}^{\alpha}_{1}}
{\bf K}_1 \hat I^{\alpha}[2,...,N] \tilde \rho_{N} =
\frac{\partial ({\bf K}_1 \tilde \rho_{1})}{\partial {\bf q}^{\alpha}_{1}} .\]

The force ${\bf F}_{1}$ acts on the first particle.
This force is a sum of the internal forces
${\bf F}_{1k}={\bf F}({\bf q}_{1},{\bf p}_{1},{\bf q}_{k},{\bf p}_{k},t)$, 
and the external force
${\bf F}^{e}_1={\bf F}^{e}({\bf q}_{1},{\bf p}_{1},t)$.
In the case of binary interactions, we have
\be \label{Fie2}
{\bf F}_{1}={\bf F}^{e}_1+\sum^{N}_{k=2} {\bf F}_{1k}. \ee
Using Eq. (\ref{Fie2}), the second term in Eq. (\ref{r1i2})
can be rewritten in the form
\[ \hat I^{\alpha}[2,...,N] \Bigl(
\frac{\partial ({\bf F}_1 \tilde \rho_{N})}{\partial {\bf p}^{\alpha}_{1}}  
\Bigr) =\hat I^{\alpha}[2,...,N] \Bigl(
\frac{\partial ({\bf F}^{e}_1 \tilde \rho_{N})}{\partial {\bf p}^{\alpha}_{1}} +
\sum^{N}_{k=2}
\frac{\partial ({\bf F}_{1k} \tilde \rho_{N} )}{\partial {\bf p}^{\alpha}_{1}}
\Bigr) = \]
\be  \label{rr1i3} =
\frac{\partial ({\bf F}^{e}_1 \tilde \rho_{1})}{\partial {\bf p}^{\alpha}_{1}} 
+\sum^{N}_{k=2}
\frac{\partial}{\partial {\bf p}^{\alpha}_{1}} \hat I^{\alpha}[2,...,N]
\Bigl( {\bf F}_{1k} \tilde \rho_{N} \Bigr) . \ee
We assume that distribution function is invariant under the
permutations of identical particles \cite{Bog2}:
\[ \rho_N(...,{\bf q}_{k},{\bf p}_{k},...,{\bf q}_{l},{\bf p}_{l},...,t)=
\rho_N(...,{\bf q}_{l},{\bf p}_{l},...,{\bf q}_{k},{\bf p}_{k},...,t) . \]
In this case, the N-particle distribution function $\tilde \rho_{N}$
is a symmetric function for the identical particles and
we have that all $(N-1)$ terms of sum (\ref{rr1i3}) are identical.
Therefore the sum can be replaced by one term with the
multiplier $(N-1)$:
\be \label{rhs} \sum^{N}_{k=2}  \hat I^{\alpha}[2,...,N] \
\frac{\partial}{\partial {\bf p}^{\alpha}_{1s}} 
\Bigl( {\bf F}_{1k} \tilde \rho_{N} \Bigr) 
= (N-1)  \hat I^{\alpha}[2,...,N] \
\frac{\partial}{\partial {\bf p}^{\alpha}_{1}}
\Bigl( {\bf F}_{12} \tilde \rho_{N} \Bigr)  . \ee
Using $\hat I^{\alpha}[2,...,N]=\hat I^{\alpha}[2]\hat I^{\alpha}[3,...,N]$, 
we can rewrite the right hand side of Eq. (\ref{rhs}) in the form
\be \label{Eq77} \hat I^{\alpha}[2] \
\frac{\partial}{\partial {\bf p}^{\alpha}_{1}} 
\Bigl( {\bf F}_{12} \hat I^{\alpha}[3,...,N] \tilde \rho_{N} \Bigr) 
=\frac{\partial}{\partial {\bf p}^{\alpha}_{1}}
\hat I^{\alpha}[2] \ \Bigl( {\bf F}_{12} \tilde \rho_{2} \Bigr) . \ee
Here, we use definition of 2-particle distribution function $\tilde \rho_2$. 
This distribution is defined by the fractional integration
 of the $N$-particle distribution function over all 
${\bf q}_{k}$ and ${\bf p}_{k}$, except $k=1,2$:
\be \label{p2} \tilde \rho_{2}=
\tilde \rho({\bf q}_{1},{\bf p}_{1},{\bf q}_{2},{\bf p}_{2},t)=
\hat I^{\alpha}[3,...,N] \tilde \rho_{N}({\bf q},{\bf p},t) . \ee
Since ${\bf p}_{1}$ is independent of ${\bf q}_{2}$, ${\bf p}_{2}$, 
we can change (\ref{Eq77}) the order of
the integrations and the differentiations:
\[ \hat I^{\alpha}[2] \ \frac{\partial}{\partial {\bf p}^{\alpha}_{1}}
\Bigl( {\bf F}_{12} \tilde \rho_{2} \Bigr) =
\frac{\partial}{\partial {\bf p}^{\alpha}_{1}}
\hat I^{\alpha}[2] {\bf F}_{12} \tilde \rho_{2}  . \]

Finally, we obtain the equation for one-particle reduced
distribution function
\be \label{er1-2} \frac{\partial \tilde \rho_{1}}{\partial t}+
\frac{\partial ({\bf K}_1 \tilde \rho_{1})}{\partial {\bf q}^{\alpha}_{1}}
+\frac{\partial ({\bf F}^{e}_1 \tilde \rho_{1})}{\partial {\bf p}^{\alpha}_{1}}
=I(\tilde \rho_{2}). \ee
Here $I(\tilde \rho_{2})$ is a term with 2-particle
reduced distribution function
\be \label{I2} I(\tilde \rho_{2})=
-(N-1) \frac{\partial}{\partial {\bf p}^{\alpha}_{1}} 
\hat I^{\alpha}[2] {\bf F}_{12} \tilde \rho_{2} . \ee
Therefore the fractional generalization of the first BBGKI equation 
has the form
\[ \frac{\partial \tilde \rho_{1}}{\partial t}=
{\cal L}_1 \tilde \rho_{1}+I(\tilde \rho_2) , \]
where ${\cal L}_{1}$ is 1-particle Liouville operator
\be \label{Lam1} {\cal L}_{1} \tilde \rho_{2} =-
\frac{\partial ({\bf K}_1 \tilde \rho_{2})}{\partial {\bf q}^{\alpha}_{1}}-
\frac{\partial ({\bf F}^{e}_1 \tilde \rho_{2})}{\partial {\bf p}^{\alpha}_{1}} 
. \ee
The physical meaning of the term $I(\tilde \rho_{2})$ is following.
The term $I(\tilde \rho_{2})d\mu_{\alpha}({\bf q},{\bf p})$
is a velocity of particle number change in $4m$-dimensional
elementary phase volume
$d\mu_{\alpha}({\bf q}_1,{\bf p}_2,{\bf q}_2,{\bf p}_2)$.
This change is caused by the interactions between particles.
If $\alpha=1$, then we have the first BBGKI equation for
non-Hamiltonian systems.  

Let us consider the particles as statistical independent systems.
In this case, we have
\be \label{2-12}
\tilde \rho_2({\bf q}_{1},{\bf p}_{1},{\bf q}_{2},{\bf p}_{2},t)=
\tilde \rho_1({\bf q}_{1},{\bf p}_{1},t)
\tilde \rho_1({\bf q}_{2},{\bf p}_{2},t) . \ee
Substituting (\ref{2-12}) in (\ref{I2}), we get
\[ I(\tilde \rho_{2})=
-\frac{\partial}{\partial {\bf p}^{\alpha}_{1}}
\tilde \rho_1 
\hat I^{\alpha}[2] {\bf F}_{12} \tilde \rho_1({\bf q}_{2},{\bf p}_{2},t) , \]
where $\rho_1=\rho_1({\bf q}_{1},{\bf p}_{1},t)$.
As the result, we have the effective force
\[ {\bf F}^{eff} ({\bf q}_{1},{\bf p}_{1},t)=
\hat I^{\alpha}[2] {\bf F}_{12} \tilde \rho_{1}({\bf q}_{2},{\bf p}_{2},t). \]
In this case, we can rewrite the term (\ref{I2}) in the form
\be \label{Ir2} I(\tilde \rho_{2})=
- \frac{\partial}{\partial {\bf p}^{\alpha}_{1}}
(\tilde \rho_{1} {\bf F}^{eff}) . \ee
Substituting (\ref{Ir2}) in Eq. (\ref{er1-2}), we get
\be \label{p1-1} \frac{\partial \tilde \rho_{1}}{\partial t}+
\frac{\partial ({\bf K}_1 \tilde \rho_{1})}{\partial {\bf q}^{\alpha}_{1}}
+\frac{\partial}{\partial {\bf p}^{\alpha}_{1}} \Bigl(
({\bf F}^{e}_1+(N-1){\bf F}^{eff}) \tilde \rho_{1} \Bigr)=0 . \ee
This equation is a closed equation for 1-particle
distribution function with the external force ${\bf F}^{e}_1$
and the effective force ${\bf F}^{eff}$.
Equation (\ref{p1-1}) is a fractional generalization of the 
Vlasov equation.

Let us differentiate this equation (\ref{p2}) that defines 
the two-particle reduced distribution function $\tilde \rho_2$. 
The fractional Liouville equation allows us to derive equation for
2-particle reduced distribution function $\tilde \rho_{2}$ in the form
\be \label{er1-4} \frac{\partial \tilde \rho_{2}}{\partial t}=
{\cal L}_{1} \tilde \rho_{2}+{\cal L}_{2} \tilde \rho_{2}+
{\cal L}_{12} \tilde \rho_{2}+ I(\tilde \rho_{3}), \ee
where ${\cal L}_{1}$ and ${\cal L}_{2}$ are 1-particle Liouville 
operators (\ref{Lam1})
and ${\cal L}_{12}$ is 2-particle Liouville operator
that is defined by equation
\[ {\cal L}_{12} \tilde \rho_{2}=
\frac{\partial}{\partial {\bf p}^{\alpha}_{1}}
\Bigl( {\bf F}_{12} \tilde \rho_{2} \Bigr)+
\frac{\partial}{\partial {\bf p}^{\alpha}_{2}}
\Bigl( {\bf F}_{21} \tilde \rho_{2} \Bigr), \]
and $I(\tilde \rho_3)$ is a term with the 3-particle
reduced distribution
\be I(\tilde \rho_{3})=  \frac{(N-1)(N-2)}{2}
\hat I^{\alpha}[3] \
\Bigl( \frac{\partial ({\bf F}_{13} \tilde 
\rho_{3})}{\partial {\bf p}^{\alpha}_{1}}+
\frac{\partial ({\bf F}_{23} \tilde \rho_{3})}{\partial {\bf p}^{\alpha}_{2}}
\Bigr). \ee
The derivation of Eq. (\ref{er1-4})
is the analogous to the derivation of Eq. (\ref{er1-2}).
It is easy to see that the system of Eqs. (\ref{er1-2}) and (\ref{er1-4})
are not closed. The system of these equations for the reduced distribution
functions can be called the fractional generalization of the BBGKI equations.


\section{Conclusion}

In this paper, we consider the fractional generalizations of Liouville 
and BBGKI equations. 
The normalization condition, phase volume, and average values are 
generalized for fractional case.
These generalizations lead us to the fractional analog of phase space. 
The space can be considered as a fractal dimensional space. 
The physical interpretation of the fractional phase space 
is discussed.
The fractional generalization of average values is derived.

In this paper the fractional analogs of the BBGKI equations are derived.
In order to derive these analogs we use
the fractional Liouville equation \cite{chaos},
we define the fractional average values and the fractional
reduced distribution functions.
The fractional analog of Vlasov equation is considered.

The fractional Liouville, BBGKI and Vlasov equations are better
approximation than its classical analogs for the systems 
with the fractional phase spaces.
For example, the systems that live on some fractals 
can be described by the suggested fractional equations.
Note that the fractional system is non-Hamiltonian systems.
The fractional harmonic oscillator
is an oscillator in the fractional phase space that can be 
considered as a fractal medium.
Therefore the fractional oscillator can be interpreted 
as an elementary excitation of some fractal medium 
with non-integer mass dimension.

It is not hard to prove that the fractional systems 
are connected with the non-Gaussian statistics.
That the dissipative and non-Hamiltonian systems can have
stationary states of the Hamiltonian systems \cite{Tarpre}.
Classical dissipative and non-Hamiltonian systems can have
the canonical Gibbs distribution as a solution of the stationary
Liouville equations for this dissipative system \cite{Tar-mplb}.
Using the methods \cite{Tar-mplb}, it is easy to prove
that some fractional dissipative systems can have
fractional analog of the Gibbs distribution (non-Gaussian statistic)
as a solution of the fractional Liouville equations.
Using the methods \cite{Tar-mplb}, it is easy to find
the stationary solutions of the fractional
BBGKI equations for the fractional systems.
Note that the interest in and relevance of fractional kinetic equations
is a natural consequence of the realization of the importance of
non-Gaussian statistics of many dynamical systems. There is already
a substantial literature studying such equations in one or more
space dimensions.

Note that the quantization of the fractional systems is a 
quantization of non-Hamiltonian dissipative systems. 
Using the method, which is suggested in Refs. \cite{Tarpla1,Tarmsu},
we can realize the Weyl quantization for the fractional systems.
The suggested fractional Hamilton and Liouville equations allow us
to derive the fractional generalization for the quantum systems
by methods suggested in Refs. \cite{Tarpla1,Tarmsu}.

The fractional BBGKI equations can be used to derive
the Enskog transport equations. 
The fractional analog of the hydrodynamic equations
can be derived from the first fractional BBGKI equation.
It is known that the Fokker-Planck equation can be derived
from the BBGKI equations \cite{Is}.
The fractional Zaslavsky's equation \cite{Zas2,Zas}
can be derived from the fractional generalization of the BBGKI equation.

\newpage
\section*{Acknowledgment}

I would like to thank 
Prof. G.M.  Zaslavsky for very useful discussions. \\

\noindent
It is a pleasure to thank X. Leoncini, G.M. Zaslavsky 
and the Organizing Committee of the International Workshop on
Chaotic Transport and Complexity in Fluids and Plasmas
for a very enjoyable conference and 
a very pleasant stay in Carry Le Rouet.


\end{document}